\documentstyle[12pt]{article}
\catcode`\@=11

\input epsf.tex

\@addtoreset{equation}{section}
\def\theequation{\arabic{equation}}

\def\@normalsize{\@setsize\normalsize{15pt}\xiipt\@xiipt
\abovedisplayskip 14pt plus3pt minus3pt%
\belowdisplayskip \abovedisplayskip
\abovedisplayshortskip  \z@ plus3pt%
\belowdisplayshortskip  7pt plus3.5pt minus0pt}
\def\small{\@setsize\small{13.6pt}\xipt\@xipt
\abovedisplayskip 13pt plus3pt minus3pt%
\belowdisplayskip \abovedisplayskip
\abovedisplayshortskip  \z@ plus3pt%
\belowdisplayshortskip  7pt plus3.5pt minus0pt
\def\@listi{\parsep 4.5pt plus 2pt minus 1pt
            \itemsep \parsep
            \topsep 9pt plus 3pt minus 3pt}}

\def\underline#1{\relax\ifmmode\@@underline#1\else
        $\@@underline{\hbox{#1}}$\relax\fi}
\@twosidetrue
\relax

\catcode`@=12

\catcode`\@=11

\def\section{\@startsection{section}{1}{\z@}{3.5ex plus 1ex minus
   .2ex}{2.3ex plus .2ex}{\large\bf}}

\def\ps@headings{\def\@oddfoot{}\def\@evenfoot{}
\def\@oddhead{\hbox{}\hfill
        \makebox[.5\textwidth]{\raggedright\ignorespaces --\thepage{}--
        \hfill }}
\def\@evenhead{\@oddhead}
\def\subsectionmark##1{\markboth{##1}{}}
}

\ps@headings

\catcode`\@=12

\relax

\def\figcap{\section*{Figure Captions\markboth
        {FIGURECAPTIONS}{FIGURECAPTIONS}}\list
        {Fig. \arabic{enumi}:\hfill}{\settowidth\labelwidth{Fig. 999:}
        \leftmargin\labelwidth
        \advance\leftmargin\labelsep\usecounter{enumi}}}
 \relax
\def\tablecap{\section*{Table Captions\markboth
        {TABLECAPTIONS}{TABLECAPTIONS}}\list
        {Table \arabic{enumi}:\hfill}{\settowidth\labelwidth{Table 999:}
        \leftmargin\labelwidth
        \advance\leftmargin\labelsep\usecounter{enumi}}}
 \relax
\def\reflist{\section*{References\markboth
        {REFLIST}{REFLIST}}\list
        {[\arabic{enumi}]\hfill}{\settowidth\labelwidth{[999]}
        \leftmargin\labelwidth
        \advance\leftmargin\labelsep\usecounter{enumi}}}
 \relax

\catcode`\@=11

\def\marginnote#1{}
\newcount\hour
\newcount\minute
\newtoks\amorpm
\hour=\time\divide\hour by60
\minute=\time{\multiply\hour by60 \global\advance\minute by-
\hour}
\edef\standardtime{{\ifnum\hour<12 \global\amorpm={am}%
    \else\global\amorpm={pm}\advance\hour by-12 \fi
    \ifnum\hour=0 \hour=12 \fi
    \number\hour:\ifnum\minute<100\fi\number\minute\the\amorpm}}
\edef\militarytime{\number\hour:\ifnum\minute<100\fi\number\minute}
\def\draftlabel#1{{\@bsphack\if@filesw {\let\thepage\relax
  \xdef\@gtempa{\write\@auxout{\string
    \newlabel{#1}{{\@currentlabel}{\thepage}}}}}\@gtempa
    \if@nobreak \ifvmode\nobreak\fi\fi\fi\@esphack}
     \gdef\@eqnlabel{#1}}
\def\@eqnlabel{}
\def\@vacuum{}
\def\draftmarginnote#1{\marginpar{\raggedright\scriptsize\tt#1}}
\def\draft{\oddsidemargin -.5truein
        \def\@oddfoot{\sl preliminary draft \hfil
        \rm\thepage\hfil\sl\today\quad\militarytime}
        \let\@evenfoot\@oddfoot \overfullrule 3pt
        \let\label=\draftlabel
        \let\marginnote=\draftmarginnote
   
\def\@eqnnum{(\theequation)\rlap{\kern\marginparsep\tt\@eqnlabel}%
\global\let\@eqnlabel\@vacuum}  }
\def\preprint{\twocolumn\sloppy\flushbottom\parindent 1em
        \leftmargini 2em\leftmarginv .5em\leftmarginvi .5em
        \oddsidemargin -.5in    \evensidemargin -.5in
        \columnsep 15mm \footheight 0pt
        \textwidth 250mmin      \topmargin  -.4in
        \headheight 12pt \topskip .4in
        \textheight 175mm
        \footskip 0pt
        
\def\@oddhead{\thepage\hfil\addtocounter{page}{1}\thepage}
        \let\@evenhead\@oddhead \def\@oddfoot{} \def\@evenfoot{} 
}
\def\titlepage{\@restonecolfalse\if@twocolumn\@restonecoltrue\onecolumn
     \else \newpage \fi \thispagestyle{empty}\c@page\z@
        \def\thefootnote{\fnsymbol{footnote}} }
\def\endtitlepage{\if@restonecol\twocolumn \else  \fi
        \def\thefootnote{\arabic{footnote}}
        \setcounter{footnote}{0}}  
\catcode`@=12
\relax

\def\ps@headings{\def\@oddfoot{}\def\@evenfoot{}
\def\@oddhead{\hbox{}\hfill
        \makebox[.5\textwidth]{\raggedright\ignorespaces --\thepage{}--
        \hfill }}
\def\@evenhead{\@oddhead}
\def\subsectionmark##1{\markboth{##1}{}}
}

\ps@headings

\relax

\def\firstpage#1#2#3#4#5#6{
\begin{document}
\newcommand{\newc}{\newcommand} 
\newc{\ra}{\rightarrow} 
\newc{\lra}{\leftrightarrow} 
\newc{\beq}{\begin{equation}} 
\newc{\eeq}{\end{equation}} 
\newc{\bea}{\begin{eqnarray}} 
\newc{\eea}{\end{eqnarray}}
\def\154{\frac{15}{4}}
\def\153{\frac{15}{3}}
\def\32{\frac{3}{2}}
\def\254{\frac{25}{4}}
\begin{titlepage}
\nopagebreak
\title{\begin{flushright}
        \vspace*{-0.8in}
\end{flushright}
\vfill
{#3}}
\author{\large #4 \\[1.0cm] #5}
\maketitle
\vskip -7mm     
\nopagebreak 
\begin{abstract}
{\noindent #6}
\end{abstract}
\vfill
\begin{flushleft}
\rule{16.1cm}{0.2mm}\\[-3mm]
$^{\star}${\small Research supported in part by the EEC under the 
\vspace{-4mm} TMR contract ERBFMRX-CT96-0090 and in part by the
project CSI-430C.}\\ 
June 1997
\end{flushleft}
\thispagestyle{empty}
\end{titlepage}}

\def\simlt{\stackrel{<}{{}_\sim}}
\def\simgt{\stackrel{>}{{}_\sim}}
\date{}
\firstpage{3118}{IC/95/34}
{\large\bf  Uncertainty relation 
and non-dispersive states in Finite Quantum Mechanics$^{\star}$} 
{E.G. Floratos$^{\,a,b}$ and G.K. Leontaris$^{\,c}$}
{\normalsize\sl
$^a$Institute of Nuclear Physics,  NRCS Demokritos,
{} Athens, Greece\\[-3mm]
\normalsize\sl
$^b$ Physics Deptartment, University of Iraklion, 
Crete, Greece.\\[-3mm]
\normalsize\sl
$^c$Theoretical Physics Division, Ioannina University,
GR-45110 Ioannina, Greece.}
{In this letter, we provide evidence for a classical sector of
states in the Hilbert space of Finite Quantum Mechanics (FQM). 
We construct a subset of states whose the minimum bound of position
-momentum uncertainty ( equivalent to an effective $\hbar$) vanishes.
The classical regime, contrary to standard Quantum  Mechanical
Systems of particles and fields, but also of strings and branes
appears in short distances of the order of the lattice spacing.
{}For linear quantum maps of long periods, we observe  that 
time evolution leads to fast decorrelation of the wave packets, 
phenomenon similar to the behavior of wave packets in t' Hooft and 
Susskind  holographic picture. Moreoever, we construct explicitly 
a non - dispersive basis of states in accordance with  t' Hooft's 
arguments about the deterministic behavior of FQM.}

\newpage

 Studies of quantum field and string theories
 in the vicinity of the horizon of black holes
 suggest the existence of a minimal length at
 the string scale\cite{ven,gross,mag,rev,kempf} and consistency with the
  Bekenstein entropy bounds for black holes\cite{Bec} requires
 the finite dimensionality of the Hilbert space of states.
 These ideas  lead t' Hooft\cite{t,t1} some years ago,
 and subsequently Susskind\cite{s} 
 to propose the holographic picture. According to 
 this picture there is a description of the physical world in
 terms of finite number of Bits of information per Planck unit 
 of area of a two  dimensional screen at the boundaries of the world.
 The particles moving in space-time are represented as two dimensional 
 areas where the  number of Bits distributed give information 
 about mass, momentum and quantum numbers of the particles.
 The black holes are represented by maximum 
information density and the corresponding number of Bits is proportional to
their mass.
 The various interactions of particles are represented by splitting
  and joining the representative two dimensional regions.
 In this picture, it is very natural to represent strings by
 strings of Bits with length and energy
 proportional to their number\cite{s,Thorn}.
 Ideas of information processing by black holes
   have been introduced many years ago by Wheeler.

Although we are far at the moment from a fundamental
 theory which encompasses
naturally  such a picture, recent findings in studies of D-branes,
\cite{brane,dvv,jm} concerning microscopic derivations of the Bekenstein 
entropy formulae and very recent works
on a specific supersymmetric matrix model approximation of 
M-theory\cite{ssb}, provide hints that the holographic
  picture is on the  right track. 
The finite dimensionality of Hilbert space in the holographic picture
suggests the existence of a  number - theoretic structure with 
discrete  space - time and  dynamical variables  related to the
Bits of information.
We remind the reader that the string theory has a number - theoretic 
(p - adic) nature which is far from being understood\cite{Fro}.

Sometime ago, a discrete version of Quantum Mechanics, 
called Finite Quantum Mechanics (FQM), using
the discrete and finite representations of the Heisenberg group
appropriate for toroidal phase spaces, has been introduced by
 H. Weyl\cite{book} and subsequently studied J. Schwinger\cite{schw}.
Balian and Itzykson\cite{bi} provided explicit expressions for quantum
 maps of  prime integer dimensions, while quantization of a family of
maps from the modular group $SL(2, Z)$ has been given by 
M. Berry\cite{Berry} for every
finite dimension $N$.

 In this work we study   physical properties of FQM,
dispersion of wave packets and the modification  of the 
uncertainty relation coming from a sector in the Hilbert  space  
of states which looks classical.
 According to t' Hooft\cite{t,t1} this is  expected  
in any system with finite dimensional Hilbert space and it is the
consequence of the existence of  bases of states which,  although they
are not eigenstates of the unitary evolution matrix chosen,  they do not
disperse under time evolution.  For a class of unitary evolution matrices, 
we construct explicitly their eigenstates which turn out to be completely
 delocalized. We construct the t' Hooft basis of non-dispersive states and 
we show  that the evolution matrix typically decorrelates Gaussian wave packets,
 property which  is observed in the interaction of wave packets 
with  the horizon of black hole and it is  consistent with the existence
of a  maximum information density. 

We also examine the violation of  the uncertainty relation which is due 
to the finiteness of the Heisenberg group.  For particular states,
we find that  the effective Planck
constant  vanishes. Of course this does not
necessarily imply that for these particular states
the dispersion in position and momentum simultaneously
can become arbitrarily small. Furthermore,
in a different subset of states evolution in time does not
lead to dispersion in position or in momentum. Obviously,
in order to have a real classical behavior
both  characteristics must simultaneously appear.

We finally note that recently various proposals
for the modification of the uncertainty principle implied by the dynamics of
D-branes have been put forward which provide evidence that the string scale 
{\it is not} the ultimate  one and it can be probed by  scattering of
D-branes \cite{autho}. Recent progress on the matrix model
interpretation of the M-theory indicates that it is possible to
formulate a theory describing these states and their dynamics.

We review first the structure of Finite Quantum Mechanics(FQM).
The classical phase space for one degree of freedom is the discrete
 square torus $T^2$ of dimensions $N\times N$ and the classical time is
 discrete\footnote{for the  Heisenberg group of phase spaces of arbitrary
 arbitrary genus see\cite{others}.}. Such a phase space can be realized
 physically as the configuration  space of the center
of the electronic circular orbits in a transverse magnetic field
periodically closing the plane to form a torus.

The simplest classical canonical transformations are the linear ones
\beq
\left(\begin{array}{c}q_{n+1}\\p_{n+1} \end{array}\right)
 =
\left(\begin{array}{rr} a&b\\c&d \end{array}\right)
\left(\begin{array}{c}q_{n}\\p_{n} \end{array}\right)
\label{2.1}
\eeq
with $a,b,c,d$, integers modulo $ N$ (mod$\, N$),
 with determinant $ ad-bc=1 $mod$\,N$
(here we differ from Berry et al\cite{Berry, ket} who take $ad-bc=1$).
Thus the linear canonical transformations form the group $SL(2,N)$.
{}For $N = p^n$ where $p$ is a prime integer, the quantum mechanical
 representation has been studied by\cite{floathn}.
{}For general $N= \prod_{i=1}^kp_i^{n_i}$ it is known
 that $SL(2,N) = \prod_{i=1}^k\,SL(2,p_i^{n_i})$  and the 
quantization is reduced with tensor products to the previous case. 

The classical harmonic oscillator corresponds to the mapping
 $a=d=0$ and $-b = c =1$
but this mapping commutes  with elements of
$SL(2,N)$ of the form $$ \left(\begin{array}{rr} 
a&b\\-b&a \end{array}\right),\,\, a^2+b^2=1 {\rm mod}\,N$$
These elements form a subgroup of $SL(2,N)$ which we call $O_2(N)$.
{}For $N$ power of a prime, this is a cyclic group
and we call the corresponding generator  the Balian - Itzykson 
(BI) oscillator\cite{bi,floath, floathn}.
 When $N=\prod_{i=1}^kp_i^{n_i}$ the group is the product of $k$
cyclic abelian groups. 

 On the $N$ dimensional  Hilbert space 
the basic operators are the shift and clock matrices
 $Q_{ij}= \omega^{s+1-i}\delta_{ij}$ and
$P_{ij}=\delta_{s+1-i,j}$ where $\delta_{ij}$ is the Kronecker
delta with indices mod($N$) and $N= 2 s +1$ where $\omega$ is the $N$-th
root of unity, $\omega = e^{\frac{2\pi\imath}N}$. 
These matrices satisfy the relation
\beq
QP = \omega PQ
\label{HCR}
\eeq
and generate
the discrete and finite Heisenberg-Weyl group $HW(N)$\cite{schw}
\beq
J_{rs}=\omega^{rs/2}P^{r}Q^{s}\qquad r,s = \{1,\ldots N\}.
\label{Hei}
\eeq
The matrices $Q$ and $P$ are connected by the finite Fourier transform
\beq
F_{kl}=\frac 1{\sqrt{N}}\omega^{(s+1-k)(s+1-l)}\label{ftr}
\eeq
with the properties,  $QF = FP$ and $F^4=1$.
The monomials  $J_{rs}$  satisfy the relations 
 \beq 
J_{rs} J_{r's'}= \omega^{(r's-s'r)/2} J_{r's'}J_{rs}
\eeq
with $J_{rs}^N=I$, and so they form a complex basis for $ SU(N)$\cite{FZ}.
One can show\cite{schw} that $HW(N)= \prod_{i=1}^k HW(p_i^{n_i})$
where the prime factors belong to the factorization of $N$.
In FQM the time is discrete and we cannot define unambiguously 
the Hamiltonian of dynamical systems but only the unitary evolution
operator of a unit time step. In the case of linear maps, i.e.
 elements of $SL(2,N)$, 
the dynamical equation which replaces the Heisenberg equation of motion
is provided by the metaplectic representation of $ SL(2,N)$\cite{bi,floath}
$$U^\dagger (A)J_{rs}U(A)=J_{r's'}$$
where $(r',s') = (r,s) A$. 
Explicit construction of the evolution matrix $U(A)$ is given in 
\cite{floath} for a matrix $A$ of $SL(2,N)$ and $N$ a prime integer:
\beq
U(A)_{kl} = \frac{1}{\sqrt{N}} (-2 c|N)
\left\{\begin{array}{c}1\\-\imath\end{array}\right\}
\omega^{-[a (k-1)^2 + d (l-1)^2 - 2 (k-1) (l-1)]/(2 c)}
\label{3.1}
\eeq
where \beq
A = 
\left(\begin{array}{rr} a&b\\c&d \end{array}\right),  
\eeq
 and $(a|p) = \pm 1 $ depending on whether
$a$ is  (or is not) the square of an integer mod$\,p$ and 
the upper (lower) value, 1($-\imath$), 
in the curly bracket corresponds to $N=4k\pm 1$.
 The exponent has to be
calculated mod$\,N$ so the inverse $1/(2 c)$ has to exist mod$\, N$.

Now we come to the implications of FQM.
The basic features of quantum mechanical behavior, is the uncertainty
relation, the dispersion of the wave-packets and the superposition
principle. Any violation of the above features is considered as violation
 of Quantum Mechanics. Field Theories share the same
characteristics plus the transformation of energy into matter and
 vice versa subject to Lorenz invariance and conservation of
Quantum numbers. The first departure from
Quantum Mechanical particle behavior is observed in string theory due 
to the non-locality of the string\cite{KKS}. The well known correction
of the uncertainty relation found in refs\cite{ven,gross,mag,kempf} in the high
energy scattering of strings for fixed impact parameter has the form
\beq
\Delta x\Delta p \ge \hbar (1+ \alpha'(\Delta p)^2)
\eeq
and implies a minimum distance of the order of the string scale
$\ell_s= g/\sqrt{\alpha'}$.
Another equivalent form of the same fact has been proposed by Yoneya
\cite{yon} which is more intuitive 
\beq
\Delta x\Delta t \ge \ell_s^2
\eeq
In this relation, the size of the string of energy $E$ contrary to the
particle behavior, is $\Delta x \propto \ell_s^2 E$\cite{s} and the time
resolution $\Delta t \propto 1/E$. This form of the uncertainty relation,
has the benefit to apply also to the semi-classical  $D$ -
branes as recently shown by Yoneya. In the recent literature there is
an extended study of possible violation of the uncertainty relation 
coming  from string defects\cite{MNE}.

In the following, we wish to investigate the form of the violation  of the
uncertainty relation in the context of FQM. Therefore we must calculate
the commutator of the position and momentum operators.
We define the  position operator $\hat{q}$ , 
$Q=e^{\imath\hat{q}}$, and so $\hat{q}$ is defined modulo diagonal integer
matrices. We choose
\beq
\hat{q}_{ij}= \frac{2\pi}{N}(s+1-i)\delta_{ij}
\label{qhat}
\eeq
By Fourier transform, the corresponding momentum operator is found to be
$P = e^{-\imath \hat{p}}$, with
\beq
\hat{p}_{ij}=-\imath\frac{\pi}N\frac{(-1)^{(i-j)}}{sin \frac{\pi}N (i-j)}
\label{mom}
\eeq
when $i\not= j$ and zero when $i=j$.

The position and momentum operators chosen have definite properties
under the parity operator $S=F^2$, i.e., $S\hat{q}S= - \hat{q}$ and 
the same holds for $\hat{p}$, as it follows from $\hat{p}=-F^{-1}
\hat{q}F$. The matrix $\hat{p}$ is a discrete version of the
 derivative of the $\delta$-function.

Incidentally, we note here that for continuous time, one
particle hamiltonians can be written as $N\times N$ matrices
\beq
{\cal H} = \frac{\hat{p}^2}{2 m}+ {\cal V}(\hat{q})
\label{hamilt}
\eeq
which can be easily treated numerically with fast convergence
to the continuum standard Quantum Mechanical results\cite{vvv}.
 We have
checked the case of harmonic oscillator for the ground state
as well as for excited states. The eigenvalues come very close
to the exact ones (exept the end of the spectrum) with 
an effective $\hbar$
being equal to  $2\pi/N$. We hope to come back in the future on 
this point.

Having defined explicitly the form of the position and momentum
operators in FQM, we calculate the uncertainty
relation for these operators
($\hat{q},\hat{p}$),
\beq
\Delta\hat{q}\Delta\hat{p}\ge \frac{1}{2}\mid <[\hat{q},\hat{p}]>\mid
\label{unre}
\eeq
where the dispersion of an observable  $A$ is defined as $\Delta{A}=
\sqrt{<({A}-<{A}>)^2>}$. 

The  commutator on the right
hand side(RHS) can be explicitely calculated 
\beq 
-\imath [\hat{q},\hat{p}]_{ij}= \frac{2\pi}N
 \frac{  \frac{\pi}N (i-j) (-1)^{(i-j)}}{
\sin{\frac{\pi}N (i-j)}}\equiv \hat C_{ij}
\label{com}
\eeq
when $i\not= j$ and zero when $i=j$. In ref\cite{ST} it
has been shown that the above commutator has the correct continuum
limit. It will prove important to know the eigenvalues of this
hermitian matrix. Although it is difficult to find explicit 
expressions for eigenvectors and eigenvalues, we observe that the 
commutator matrix $\hat{C}$ for fixed $(i-j)$ goes in the large $N$ 
limit to the matrix $\hat{C}_{\infty}$
\beq
\hat{C}_{ij}\ra \hat{C}_{\infty}\equiv \frac{2\pi}N (-1)^{(i-j)}
\delta_{ij}\label{clim}
\eeq
The above matrix has eigenvectors 
\beq
v_k=\left\{\frac 1{\sqrt{N}}
 (-1)^l\omega^{l\cdot k}\right\}_{l:0,\ldots , N-1}
\label{veclim}
\eeq
and eigenvalues, $-\frac{2\pi}{N}$ for $k=1,\ldots , N-1$ and
$\frac{2\pi}{N}(N-1)$ for $k=0$. Except from the last eigenvalue,
this reminds the situation in the
standard quantum mechanics if we identify $\hbar =\frac{2\pi}N$.
For finite $N$, the matrix $\hat{C}$ commutes with the finite
{}Fourier transform so they share common eigenvectors. Numerical 
investigation shows that for relatively large $N\sim 100$,
there are few big negative eigenvalues and most of the remaining
eigevalues are very close to $2\pi/N$.

We discuss now the implications of the new uncertainty relation
(\ref{unre}). For a general wave function
\beq
\psi = \sum_{i=1}^Nc_i\psi_i,\;\; \sum_{i=1}^N |c_i|^2 =1,
\eeq
where $\psi_i$ are the normalised eigenvectors of the hermitian
commutator matrix $\hat{C}$,
\beq
\hat{C}\psi_i = \epsilon_i \psi_i
\eeq
we find that
 \beq
 \Delta\hat{q}\Delta\hat{p}\ge 
 \frac 12\mid \sum_{i=1}^N|c_i|^2\epsilon_i\mid
\equiv \frac 12\hbar(\psi)\label{unre1}
 \eeq
It would be desirable to write the RHS of (\ref{unre1}), which is an 
effective Planck constant as a function of $\Delta\hat{q}$
and $\Delta\hat{p}$, but this is not in general possible. 
{}For the limiting matrix $\hat{C}_{\infty}$
\beq
\hbar(\psi) = \frac{2\pi}N\mid 1-N|c_N|^2\mid\label{hbarinf}
\eeq
Because some of the
$\epsilon_i$'s have to be negative, we can find an $(N-2)$-
dimensional subset ${\cal H}_{cl}$ of the $N$-dimensional complex 
unit sphere for which  $\hbar(\psi)$ vanishes. 
To get some feeling of the geometry of ${\cal H}_{cl}$ we observe that
the latter is invariant under the group of complex matrices which preserve
$\hat{C}$, i.e., $U^{\dagger}\hat{C}U = \hat{C}$. 
Such an example is the Finite Fourier Transform matrix. Defining a non-
positive definite inner product 
\beq
(\psi_1,\psi_2) = <\psi_1, \hat{C}\psi_2>
\label{inner}
\eeq
we find that if $\psi_{1,2}$ belong to ${\cal H}_{cl}$ and moreover
they are orthogonal with respect to this inner product, then any linear
combination of them belongs to ${\cal H}_{cl}$. The set of matrices which
preserve $\hat{C}$ form a group which is isomorphic to the non-compact
group $U(n,m)$ where $n(m)$ are the numbers of positive (negative)
eigenvalues of $\hat{C}$. On the other hand, in order to stay on the
unit complex $N$- dimensional sphere, we have to restrict to the 
diagonal compact subgroup $U(n)\oplus  U(m)$.
 
Numerical study of the $\hat{C}$ eigenstates shows smooth localization
 structure indicating possible analytic expression for their exact forms.
In the figures (1,2) we plot some eigenstates of the commutator $\hat C$
for $N=47$ which correspond to non-degenerate eigenvalues. The parity
symmetry is realized in all eigenstates and we observe that the simplest
structure appears for the most negative eigenvalue. Also, plotted is
the eigenvector corresponding to the smallest positive eigenvalue
which happens to be smaller than $2\pi/N$.

\begin{figure}
\begin{center}
\leavevmode
\epsfysize=3in \epsfbox{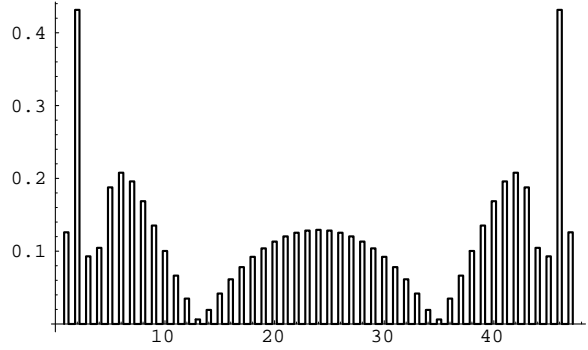}
\caption[]{ 
The eigenvector of the commutator for $N=47$ corresponding to
the largest negative eigenvalue. }
\end{center}
\end{figure}
The commutation relation (\ref{unre}) breaks explicitly translation
invariance since the shift operator $P$ does not commute with $\hat{C}$. 
The violation of translation invariance
 implies a non-trivial dependence of the RHS of the
uncertainty relation on the momentum spectrum of the wavefunction
$\psi$. 
One possible way to probe the  dependence on the width
of the momentum spectrum is to saturate the RHS for a class of
 Gaussians $\psi_G$ of various widths. 

In figure 3 we plot the RHS of the uncertainty relation (\ref{unre1})
for Gaussians of various widths in units of lattice spacings. We observe
that for widths bigger than a few lattice spacings the RHS becomes 
$2\pi/N$ which is the equivalent of $\hbar$ for FQM. Going to 
smaller widths we find a sharp transition to classical behavior
because $\hbar(\psi_G)$ goes to zero. Thus, the classical regime appears
in the order of the lattice spacing where the FQM exhibits
deviation from the standard  uncertainty relation.
 The classical sector ${\cal H}_{cl}$, must disappear in the appropriate
limit where the standard quantum mechanics is recovered, but for every
finite $N$ it signals a completely different behavior from particle
string or $D$ - brane uncertainty relation.

\begin{figure}
\begin{center}
\leavevmode
\epsfysize=3in \epsfbox{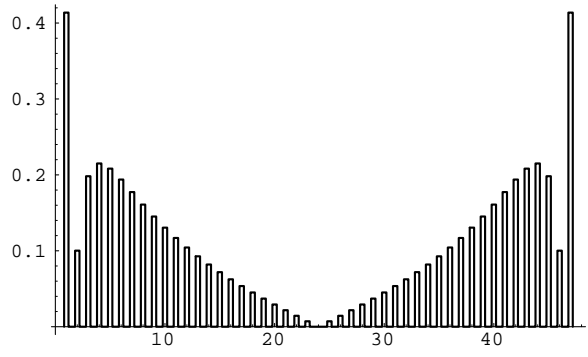}
\caption[]{ The eigenvector of the commutator for $N=47$ corresponding to
the smallest positive eigenvalue.
 }
\end{center}
\end{figure}
\begin{figure}
\begin{center}
\leavevmode
\epsfysize=3in \epsfbox{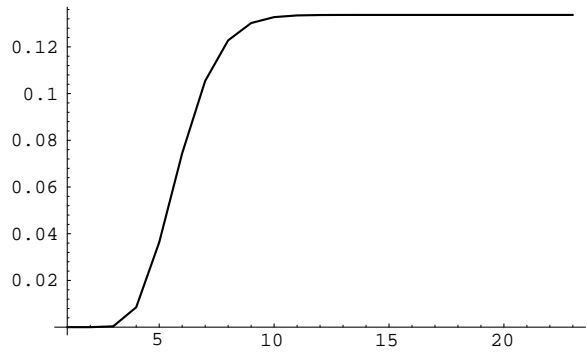}
\caption[]{ $\hbar$ for Gaussian widths  from 1 to 4 lattice spacings for
the case  $N=47$.  }
 \end{center}
\end{figure}
 

We now construct the t' Hooft non - dispersive states
for a class of specific evolution  matrices of FQM.
The BI harmonic oscillator group of matrices 
\beq
A = 
\left(\begin{array}{rr} a&b\\-b&a \end{array}\right)
\eeq 
$a^2+b^2=1$ mod$\,N$
is cyclic abelian group with $4 k$ elements and its generator
$R_0 = \left(\begin{array}{rr} a_0&b_0\\-b_0&a_0 \end{array}\right)$
for primes up to $20,000$ were determined by a search algorithm
in ref\cite{floath}.
{}For $N= 4 k +1$, $R_0$ is diagonalized by a matrix in $SL(2,N)$
\beq
R_0 = T\left(\begin{array}{rr} 
a_0-tb_0&0\\0&a_0+tb_0 \end{array}\right)T^{-1}
\label{3.3a}
\eeq
\bea
T=\frac 1{1+t}\left(\begin{array}{rr} 1&1\\-t&t \end{array}\right)
\label{3.3b}
\eea
with $t=g^k$ mod$\,N$. 
Since the $4 k$ - period of $R_0$ and $g$ coincide we see that
$a_0-t b_0$ turns out to be a primitive element $g$. Thus it is 
simple to construct  the generator $R_0$, 
\beq
R_0 = \left(\begin{array}{rr} 
\frac{g+g^{-1}}2&\frac{g^{-1}-g}{2t}\\
\frac{g-g^{-1}}{2t}&\frac{g+g^{-1}}2\end{array}\right)
\label{3.4}
\eeq
Now, since $U(A)$  defines a representation of
$SL(2,N)$, we have,
\beq
U(R_0) = U(T) U \left(\begin{array}{rr} g&0\\0&g^{-1} \end{array}\right)
          U^{-1}(T)
\label{3.5}
\eeq
The eigenvalue problem for $U(R_0)$   was solved analytically
for all primes of the form $N= 4 k +1$
in ref\cite{floath,floathn}. The observation is simply that 
$U\left(\begin{array}{rr} g&0\\0&g^{-1} \end{array}\right)$ is by
construction the matrix
\beq
D_{k,\ell} = - \delta_{{k-1},g(\ell-1)}\,\,\;\;
\, k,\ell = 1,\ldots , N
\label{3.8}
\eeq
which has eigenvalues 
$
- e^{\frac{\pi\imath}{2k}\ell}, \ell = 1,...,4 k
$ and $-1$.
The eigenvectors are determined by  the multiplicative characters
of the finite field $F_p = \{ 0,1,\ldots , p-1\}$ and these are
the ($4k+1$)-dimensional vectors $\Pi_j=\{0, \Pi_j(g^n)\}$
where 
\beq
\Pi_j(g^n) = \frac{e^{\frac{\imath\pi}{2k}jn}}{\sqrt{4k}},\,\,
\, j,n =1,2,\ldots , 4 k
\label{3.9}
\eeq
whilst the $(4 k +1)-{th}$ eigenvector is defined to be $\{1,0,...,0\}$.\\
To proceed, we observe that the circulant matrix ${D}$, in (\ref{3.8})
permutes the axes vectors $e^i$,
\beq
e^i_j = \delta_{ij}\label{3.12a}
\eeq
as follows
\beq
D^m e^i = (-)^m e^{1+g^m(i-1)}\label{3.13}
\eeq
So the non-dispersive t' Hooft states for $G=U(R_0) =
U(T) C U^{-1}(T)$ in (\ref{3.5}) are
$
u^i = U(T) e^i
$
and they evolve as
\beq
D^m u^i = (-)^m u^{1+g^m(i-1)}.
\label{3.15}
\eeq
Using (\ref{3.1}) we find explicitely the $U(T)$ and $u^i$'s
are found to be
\bea
(u^i)_k =
 \frac 1{\sqrt{N}} \left(((1+t)|N)\right)
 \omega^{\frac{1}{2}(t(k-1)^2+(i-1)^2+2(1-t)(k-1)(i-1))}
 \label{3.16}
 \eea
We observe that all the components of $u^i$'s are pure phases.
The evolution law given by (\ref{3.15}) implies that the dispersion of 
the position operator in the states $u^i$,
\bea
\left(\Delta\hat{q}\right)^2_i = 
{u^i}^\dagger\left(\hat{q}-<\hat{q}>_i\right)^2u^i,\,&
<\hat{q}>_i= {u^i}^\dagger\hat{q}u^i
\label{3.17}
\eea
remains constant, 
since all $(u^i)_{\ell}$ are pure phases and they remain so under
evolution with $U(R_0)$.
The same happens to the dispersion of momentum operator because
$U(R_0)$ commutes with the finite Fourier transform, 
$F = \imath^k U(R_0)^k$.

According to t' Hooft, the set of the non-dispersive states is
not observable by localized experiments but only the effects of
linear combinations of a big number of them which produce 
localized states (particles). In terms of the non-dispersive
basis, FQM is a deterministic theory. For the quantum linear
maps the non- dispersive states permute among themselves during 
evolution in a random way. This implies that Gaussian wave packets
seen as linear combinations of the non-dispersive states basis
will decorrelate quickly and become delocalized. This can be
understood by the following semi-classical argument: Since 
linear maps of long period produce long random trajectories and 
the corresponding wavefunctions have to be extended along them, 
Gaussian wave packets at time $t_0$ will disperse under time 
evolution along random the trajectories\cite{floathn,viv}.

This property is
reminiscent of the behavior of wavepackets falling on the
horizon of a black hole which become extended along the
horizon due to a minimum information density per Planck
area. This is also true for strings which become extended
and are wrapped around the horizon\cite{t,s}.

We coclude our discussion with the following observations.
The classical sector of FQM is characterized by the following
two features:\\
1) There is a subset of states $\psi_i$  where the effective Planck
constant $\hbar(\psi_i)$ vanishes. This however, does not 
necessarily imply that for the particular (Gaussian) states  
the dispersion in position and momentum simultaneously
can become arbitrarily small. In fact the product of the
position momentum uncertainty is bounded
to be bigger than $1/16\pi^2$ due to the property of
the Fourier transform. The intringuing property here, 
is the vanishing of the commutator
for very narrow Gaussians.\\
2)In a different subset of states evolution in time does not
lead to dispersion in position or in momentum.\\
The above two characteristics must simultaneously appear in order
to have a real classical behavior. In other words, one should determine
the intersection of the two different subsets of the Hilbert space.
This problem remains  open for future investigation.

Another novel feature of FQM is the existence of quantum maps
which decorrelate  Gaussian initial states and distribute
the wave packet information equally among the points of the
configuration space. This is a desirable property according to
the holographic picture which preserving unitarity randomizes
the initial information in a retrievable way.

{{\bf Acknowledgements:}{\it \, We would like to thank CERN theory division
for kind hospitality while parts of this work has been done. We also
thank Gregory Athanasiu and A. Polychronakos for interesting discussions.}} 
 

\begin{thebibliography}{99}  
\bibitem{ven} G. Veneziano, Europhys. Lett. {\bf 2}(1986) 199;
D. Amati, M. Ciafaloni and G. Veneziano, Phys. Lett. {B216}(1989)41.
\bibitem{gross}D.J. Gross and P.F. Mende, 
Nucl. Phys. {\bf B303}(1988)407.
\bibitem{mag}K. Konishi, G. Paffuti and P. Provero,
Phys. Lett. {\bf B 234}(1990) 276;
M. Maggiore, Phys. Lett. {\bf B304}(1993) 65.
\bibitem{rev}L. J. Garay, Int. Jour. Mod. Phys. {\bf A 10}(1995)145.
\bibitem{kempf}A. Kempf and G. Mangano, Phys.Rev.{\bf D55}(1997)7090.
\bibitem{Bec}J. Bekenstein, Phys. Rev.{\bf D 30}(1984)1669;
Phys. Rev.{\bf D 49}(1994)1912.
\bibitem{t}G. t' Hooft, Nucl. Phys. {\bf B 342} (1990) 471.
\bibitem{t1} G. t' Hooft, Dimensional Reduction in Quantum Gravity,
{\tt gr-qc/9310006}.
\bibitem{s}  L. Susskind, Phys. Rev. {\bf D 49}(1994)6606;
 Journ. Math. Phys. {\bf 36}(1995) 6377.
\bibitem{Thorn}O. Bergman and C. Thorn,
Phys. Rev. {\bf D 52} (1995) 5980; {\tt hep-th/ 9702068}. 
\bibitem{brane}J. Polchinski, Phys. Rev. Lett.{\bf 75}(1995)4724;
E. Witten, Nucl. Phys. {\bf 460}(1995)335;
P.K. Townsend, Phys. Lett. {\bf B 373}(1996)68;
A. Strominger and C. Vafa, Phys. Lett.{\bf B379}(1996)99.
\bibitem{dvv}R. Dijkraff, E. Verlinde and H. Verlinde, Nucl. Phys. {\bf B 486}
(1997)77.
\bibitem{jm}J. Maldacena, Black Holes in String Theory, Princeton Univ. Thesis,
{\tt hep-th/9607235}. 
\bibitem{ssb}
T. Banks, W. Fischler, S.H. Shenker, and L. Susskind,
Phys. Rev.{\bf D 55}(1997) 5112;{\tt hep-th/9610043};
R. Dijkgraaf, E. Verlinde, H. Verlinde {\tt hep-th/9703030}.
\bibitem{Fro}L. Brekke, P. Freund, M. Olson and E. Witten,
Nucl. Phys. {\bf B 302}(1988) 365\\
L. Brekke and P. Freund, Phys. Rep. {\bf 233}(1993)1.
\bibitem{book} H. Weyl, Theory of Groups and Quantum Mechanics,
Dover, 1930.
\bibitem{schw}J. Schwinger, J. Math. Phys. {\bf 2}(1961) 407.
\bibitem{bi}R.  Balian and C. Itzykson, C. R. Acad. Sci. Paris,
{\bf 303} (1986) 773.
\bibitem{Berry}M. Berry et al, Annals Phys.{\bf 122}(1979)26-63;
J. Hannay and M. Berry, Physica {\bf D1}(1980)267. 
\bibitem{autho}M. Douglas, D. Kambat, P. Pouliot and S. Shenker,
 Nucl. Phys. {\bf B 485}(1997)85;
U. Danielsson, G. Ferreti and B. Sundborg,
{\tt hep-th/9603081}. 
\bibitem{others}A. Jaffe et al, Com. Math. Phys. {\bf 126}(1989)421.
\bibitem{ket} J. P. Keating, J. Phys. {\bf A 27}(1994)6605.
\bibitem{floathn} G. Athanasiu, E. G. Floratos and Nicolis, 
J. Phys. {\bf A }29(1996)6737,  {\tt hep-th/9509098}.
\bibitem{floath}G. Athanasiu, E. G. Floratos, Nucl. Phys. {\bf
 B 425}(1994) 343.
\bibitem{FZ}D. Fairlie, F. Fletcher and C. Zachos, Journ. of Math. Phys.
{\bf 31}(1990) 1088.
\bibitem{KKS}
M. Karliner, I. Klebanov and L. Susskind, Int. J. Mod. Phys. 
{\bf A 3} (1988) 1981.
\bibitem{yon} T. Yoneya and Miao Li, {\tt hep-th/9611072}
\bibitem{MNE}G. Amelino-Camelia,
 J. Ellis, N. Mavromatos and D.V. Nanopoulos, {\tt hep-th/9701144};
F. Lizzi and N. Mavromatos, {\tt hep-th/9611040},
Phys. Rev. {\bf D 55} (1997) 7859.
\bibitem{vvv} P. Stovicek and J. Tolar, Rep. on Math. Phys.
{\bf 20}(1984) 157.
\bibitem{ST} T. S. Santhanam and A. R. Tekumalla, 
Found. of Phys. {\bf 6} (1976) 583.
\bibitem{viv} M. Bartuccelli and F. Vivaldi,
Physica {\bf D 39}(1989)194.
\end{thebibliography}
\end{document}